# Co-Creating Educational Project Management Board Games to Enhance Student Engagement

Vasilis Gkogkidis [a*] & Nicholas Dacre [a]

[a] University of Southampton Business School, University of Southampton, Southampton, SO17 1BJ, UK
[*] Corresponding Author: v.gkogkidis@Southampton.ac.uk

**Abstract**

Management education scholarship has long outlined the need to enhance student engagement and participation in business schools, using more innovative teaching practices. This is increasingly motivating scholars to strive for more collaborative pedagogic dynamics between teachers and students. At the same time, research into co-creation of Game Based Learning material such as board games has largely focused on the value added to games when educators involve students in the design process. However there has been scant research examining the qualities of co-creational game design exercises as teaching experiences themselves, thus overlooking the opportunity to conceptualise such activities as an innovative teaching tool that can help educators facilitate student engagement and participation. To address this research gap, this paper presents a case study where Project Management students participated in two co-creation workshops designing educational Project Management games. Data were collected conducting focus groups at the end of the two workshops. Throughout the paper we have sought to present some positive outcomes of such processes as well as some critical points that emerged through the data that were collected. Mentionable outcomes include a series of positive characteristics of co-creative Game Based Learning activities like enhanced engagement as well as a list of challenges when facilitating such activities. The main findings of this research have been organised in two frameworks, one outlining five positive characteristics of co-creative Game Based Learning activities: engagement with knowledge, knowledge assessment, creativity, communication and the second outlining challenges in facilitating such activities: lack of focus, lack of structure and the need for more practice-oriented games. The suggested frameworks can assist educators conceptualise and utilise such exercises to create more effective and participatory learning environments.

# Introduction

Teaching and learning practices in business schools have been under critique for failing to offer a curriculum that prepares future managers for the challenges they will face in the workplace, resulting in alumni struggling to cope with the complexities of organisational life (Bennis and O'Toole, 2005; Dacre, Senyo, and Reynolds, 2019; Pfeffer and Fong, 2002; Taylor, Thorpe and Down, 2002; Ojiako et al., 2011). Engaging with student centred learning discourses, project management educators have also questioned conventional dynamics and assigned roles between students and educators in the classroom, where tutors are considered to encompass all knowledge and students are merely passive receivers of this knowledge (Ojiako et al., 2011).

Suggestions around educators assuming more facilitative rather than instructive roles to create learning environments where students can learn more independently (Long and Holeton, 2009) are becoming more prominent in management education literature alongside indications that students opinions and views should be taken into account by the academics teaching them (Del Corso, Ovcin and Morrone, 2005).

The need for new innovative teaching practices has been further outlined by (Holman, 2000) calling for pedagogies that include more experiential learning, reflection and critical thinking, posing that such practices can be beneficial for the educators as well, giving them the opportunity to reflect and improve their teaching practices.

The aim of this research was to examine if co-creative game-based learning activities can enhance student engagement and participation when discussing Project Management in business schools. The term Co-creative Game Based Learning (CGBL) describes activities which can help create a much more fruitful partnership between educators and students, where students and educators collaborate to create educational learning games (Bagheri, Alinezhad and Sajadi, 2019; Calderwood, 2019; Dacre, Gkogkidis and Jenkins, 2018; Kuhmonen et al., 2019). The contribution this paper seeks to make is towards conceptualising these processes as valuable learning methods, rather than means to an end, an endeavour motivated by results presented by Kuhmonen et al. (2019).

# Context

### Improving Student Engagement with Game Based Learning

Use of game elements and games in non-gaming situations offers the potential for enhanced engagement and participation in the activity that is taking place (Dacre, Constantinides and Nandhakumar, 2015; Nacke and Deterding, 2017). Promising applications of gamification and game-based learning have also been reported in higher educational contexts where such pedagogical tools enhanced student participation and academic performance (Ebner, M. and Holzinger, 2007; Fotaris et al., 2016).

The term GBL refers to learning environments where gameful learning experiences offer students challenges and problems designed to support learning and knowledge acquisition (Kirriemuir and McFarlane, 2004). Literature reviews examining the effectiveness of GBL in various educational contexts, report enhancement of student motivation, engagement with the curriculum as well as the creation of opportunities for the students to develop skills and knowledge connected to their curriculum (Abdul Jabbar and Felicia, 2015; Boyle et al., 2014; Randel et al., 1992; Vogel et al., 2006).

Results indicating that students can develop problem solving and communication skills in multisensory settings encouraging creativity and sensemaking of learning content (Ermi and Mäyrä, 2005) are especially important for management educators wanting to tackle the abovementioned issues in management education where students do not get opportunities to develop such skills. Board games specifically have long been considered valuable educational tools and have been effectively used in classrooms of different educational levels (Plass, 2020; Tsarava, Moeller and Ninaus, 2018).

**Co-Creation of Curricula**

The term co-creation refers to educators allowing students to take part in the design process of the curriculum (Bovill, 2013). Literature on Co-Creation of high education curricula concerns itself with questioning how inclusive traditional teaching practices are and researching the potential benefits of making these processes more participatory by involving students (Bovill, 2013).

Co-creation approaches aim at creating a more democratic type of education, enhancing student participation and engagement with the learning material and the students' degree subject at large, improving teaching practices and promoting a model of knowledge creation where knowledge is negotiated, and students are viewed as active stakeholders rather than customers (Bovill, Cook-Sather and Felten, 2011; Cook-Sather, A., Bovill, C., & Felten, 2014).

Co-creative GBL practices where students are invited to participate in the creation of learning games, is a practice operationalising such ideas, a practice that has been reported to result in GBL learning material better suited for the needs of students (Bagheri, Alinezhad and Sajadi, 2019; Calderwood, 2019; Dacre, Gkogkidis and Jenkins, 2018; Kuhmonen et al., 2019). This research suggests that co-creative GBL activities not only lead to better GBL material but also present salient learning opportunities for students and teachers alike.

Problem Based Learning (PBL) theories and practices are helpful to conceptualise the learning process of co- creative GBL activities. Designing teaching and learning using PBL theories involves students trying to solve complex real world problems that might not have a single correct answer (Poikela, Vuoskoski, and Kärnä, 2009), very much like the workshops facilitated during this research, where there was no one single best game design.

During PBL projects, students collaborate in groups identifying and applying the relevant knowledge needed to solve the problem at hand with teachers assuming a more facilitative rather than prescriptive role (Dolmans et al., 2001). Central characteristics of such approaches include peer, collaborative and reflective learning allowing students to achieve learning objectives set by their educators (Hmelo-Silver, 2004; Jensen and Krogh, 2013). Poikela and Poikela (2006) suggest the following framework outlining the stages students go through during a PBL project.

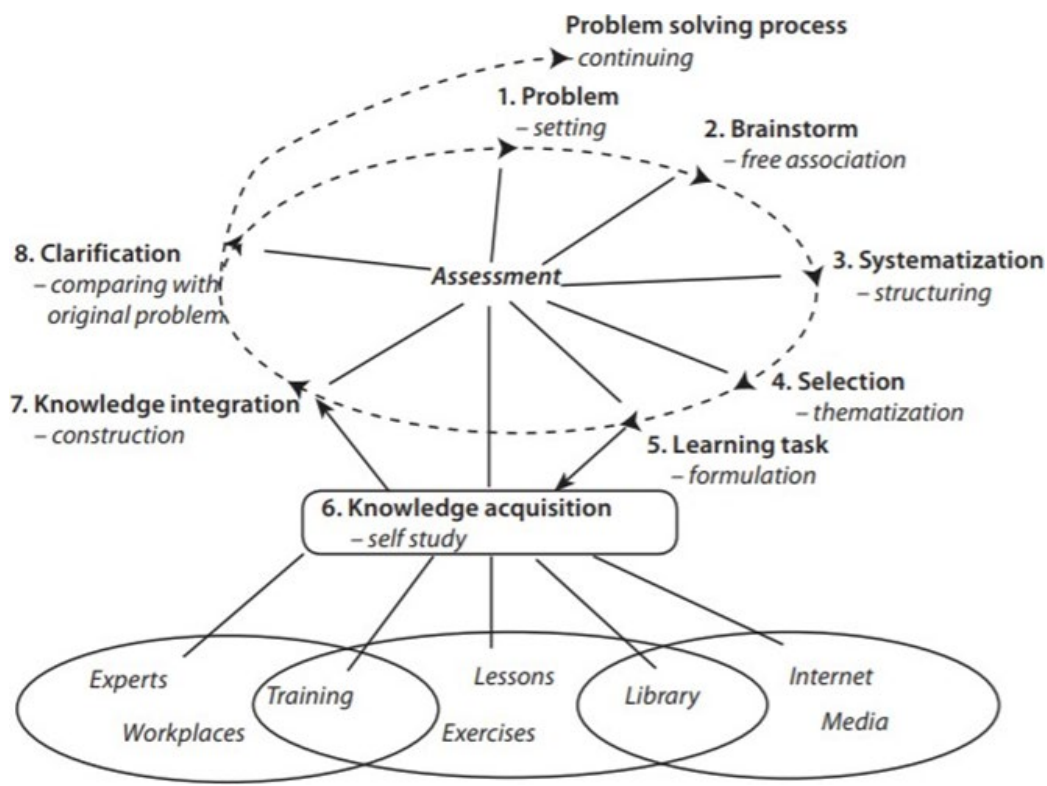

Figure 1: The problem-based learning cycle and knowledge acquisition (Poikela and Poikela, 2006)

Even though the above framework refers to student projects undertaken usually during a longer period of time, it broadly captures the journey that participants went through during the workshops designed for this research.

## Methodology

Focus groups were chosen as the data collection method because they underpin an environment where both interviewers and interviewees can share and negotiate their experiences and meaning they attach to them (Cohen, L., Manion, L., & Morrison, 2017). Focus groups allow for in-depth exploration of the issues that are being discussed, to determine the fashion in which participants shape their ideas around the current issues, including connections between these ideas and observable behaviour (Hochschild, 2009). Data were thematically analysed until the elements of the two frameworks emerged describing positive results of Co-creative GBL activities and challenges around facilitating such activities.

Two co-creation workshops were facilitated by the researchers, with each workshop lasting two hours. A total of eight Project Management undergraduate students volunteered and participated in both workshops that took place outside teaching hours. Students were randomly allocated in teams with each team designing one educational Project Management board game.

The first workshop included a short introduction to game design, alongside some relevant team exercises to get participants to reflect on games they played in the past and their characteristics. The rest of the workshop was spent with teams working on a first prototype of their board game. A number of project management frameworks were suggested as potential topics that students could pick as the central learning content of the games, however participants were free to manage their own process within their teams without any specific roles or responsibilities allocated by the researchers.

Fifty minutes were allowed for the prototyping phase during which both researchers acted as facilitators/advisors providing feedback on the range of ideas and game designs that emerged and ensuring that both teams were progressing adequately. Researchers decided to not be members of any team, to avoid power-expert dynamics where students would wait for educators to lead the design effort because 'they know more'.

The final stage of the first workshop involved playtesting both games, offering feedback and ideas for improvement. Teams were asked to explain the rules of their game and observe the other team play so that game elements that worked well and elements in need of further improvement could be identified. During the second workshop students implemented changes based on the feedback given, producing a final prototype. Finally, one focus group with each team was conducted at the end of the second workshop.

Who Wants to be a Project Manager was one of the two prototypes that were created during the workshops deploying some Who Wants to be a Millionaire like game mechanics mixed with some Snakes and Ladders like mechanics. Two teams of two players each, compete against each other to reach the end of the board.

Taking turns, each team rolls the dice to determine where they land and the other team asks a PM related question before allowing them to move to that square. One of the interesting elements of the game was that there were different types of questions available spanning a good amount of PM theory and literature.

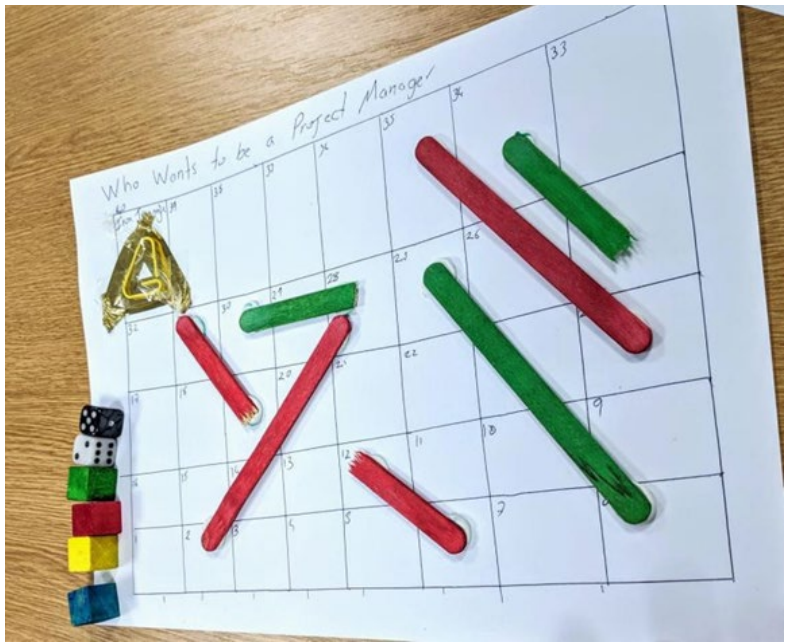

Figure 2: Who Wants to be a Project Manager, a two teams' game where participants test their PM knowledge

The second prototype created during the workshops was called The Project Adventure and it was a more straightforward game where players would compete to reach the end of the board answering questions and making some interesting choices along the way.

This game tried to incorporate a simulation feel to its gameplay as players had a budget they could spend along the way to counter various drawbacks that occurred, like in real Project Management.

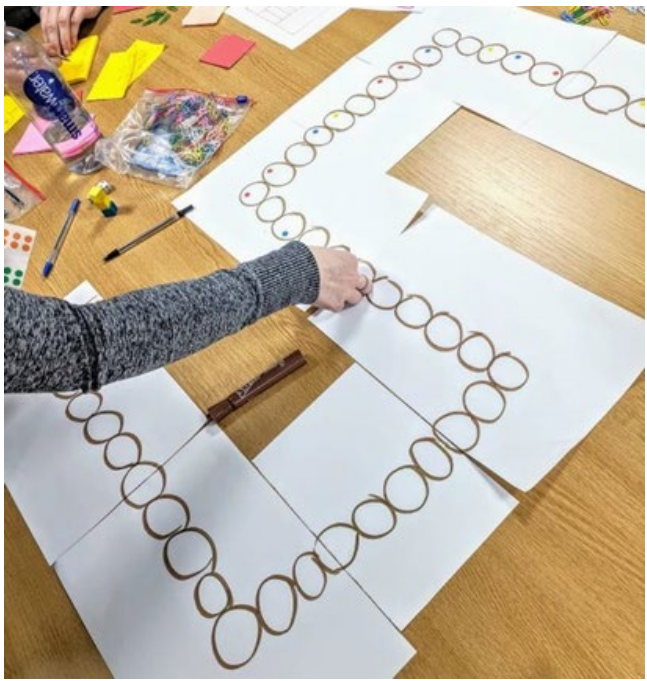

Figure 3: The Project Adventure, a competitive game where players try to reach the end of the project first

# Characteristics and Challenges of Co-Creative GBL Activities

This section outlines the five characteristics of co-creative GBL activities summarised as enhanced: engagement with knowledge, knowledge assessment, creativity, communication and collaboration as well as the challenges of facilitating such learning activities.

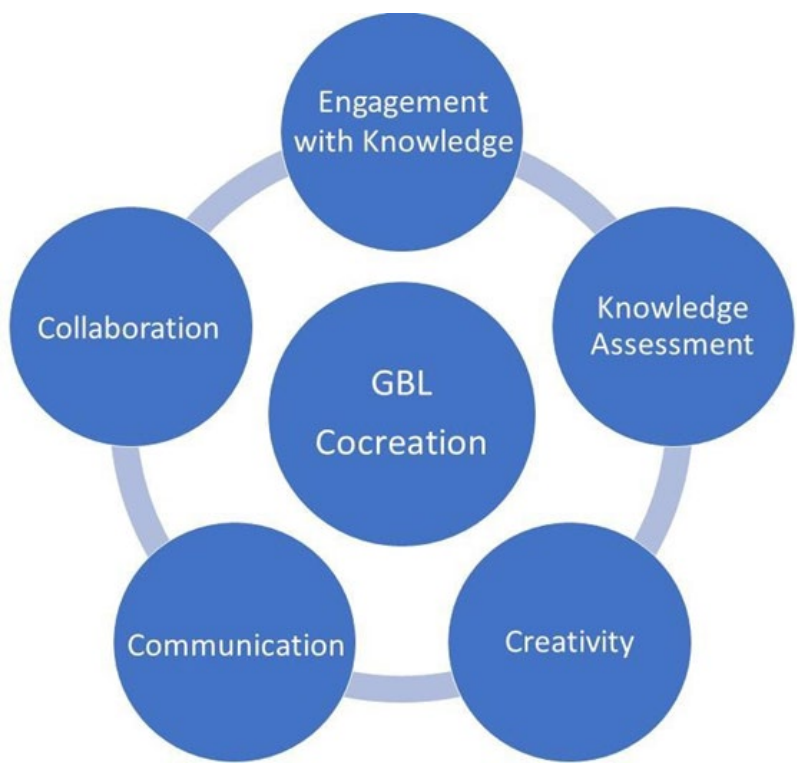

Figure 4: Characteristics of Co-Creative GBL Activities

**Engagement with Knowledge**

Engagement can be defined as an individual student's psychosocial state: their behavioural, emotional, and cognitive connection to their learning (Fredricks, Blumenfeld, and Paris, 2004). The educational challenges of student engagement and participation are considered important indicators not only of student achievement and retention (Krause and Coates, 2008) but also of the quality of a university course and it has been argued that they should be measured and factored in when trying to improve teaching methods and processes (Coates, 2005). Co-creation teaching activities have resulted in improving student engagement and participation as well as learning (Iversen et al., 2015), with similar themes of enhanced engagement with knowledge shared among participants of this research:

> "Having to create a board game based on project management, simultaneously made the workshop fun and informative as we engaged with course material from a more entertaining angle."

> "These workshops are a good way to talk about the subject area, you have to find an application for that knowledge in a game-based scenario."

The practical application of knowledge to make a real-world artefact provided the background for a lot of active learning from students saying that:

> "I truly feel that we better understand and internalise teachings when we are forced to use them in practice; as you need to fully understand something before you can constructively apply it in the creation of something else. This is something I feel a traditional teaching session is sorely lacking in."

**Knowledge Assessment**

One theme emerging from our data revolved around formative assessment where teachers and students are able to identify knowledge gaps that could be addressed in future learning sessions. Formative assessment aims at supporting student learning by offering feedback during the duration of a course (Yorke, 2003).

> "Teachers could possibly see where students have not fully understood an aspect of the subject area which would allow them to revise the area later. This process could enable teachers to assess if students actually understand how a concept fits into a greater project management picture"

Insights shared by one of the authors furthers that point by adding that such processes are useful in reflecting one's own teaching skills and approaches:

> "This process enables me to observe students potentially struggling with a piece of knowledge I take for granted, enabling me to reflect on my own teaching. Maybe I did not explain that part very well or I need to make it more explicit. We sometimes make assumptions that if there are no questions everyone understood what we said"

Through creating a learning board game, students were able to reflect and identify their own knowledge gaps, a process that can be thought of as part of self-assessment defined as a process where students self-reflect on the results of their learning efforts but do not grade themselves, self-assessment is 'feedback for oneself from oneself' (Andrade and Du, 2007, p. 60):

> "The process allowed me to identify my weakness areas as I had to come up with questions for our board game. I also identified knowledge gaps when playing the game and attempting to answer questions written by my teammates. The relaxed atmosphere of the session gave myself and my peers a chance to discover different angles of what we actually know from the module and what we thought we knew, therefore giving us a good insight in what we need to revise for"

## Creativity

(Vygotsky, 1967) established a connection between play and creativity suggesting that being immersed in play can serve as a tool for children to experiment with potential outcomes of their actions without real world implications, bringing to mind educational tools such as serious games, simulations, role-playing and board games. The goal of the workshops organised was to create such an environment where students could safely experiment with their ideas around what an educational PM game could look like.

To encourage creativity, we chose to let participants "play" with their ideas and design games as they wish, there were no specific guidelines given other than, at the end of the fifty minutes every team needed to have:

- A physical prototype that people can play
- A name for their game
- The rules of each game written on a piece of paper

> "In previous group assignments I have undertaken there has been a common theme throughout which was the abundance of guidelines. Having these strict set of criteria to adhere to hinders the creativity element of assignment process"

> "Students are more engaged because they are designing how they want to be taught and how they want to learn especially for the more creative in the group it would be a more rewarding experience allowing them to learn quicker and better that way as opposed to being told what to do"

Allowing students to approach the challenge of building an educational game any way they wanted created a feeling of ownership over the end results of the workshops as participants made efforts to produce the best game they could in the time given. Some negative feedback about the freedom we offered students can be found in the section presenting challenges around facilitating co-creative workshops.

## Communication

Themes of classroom estrangement and alienation had a salient presence during conversations with workshop participants. Both students and teachers posited alienation as being an intrinsic characteristic of traditional transmissive teaching techniques like lecturing and even seminars to a certain extend.

> "Most of the time, traditional teaching is a one-way communication between lecturers and students where the lecturers talk whilst the students listen. Thus, it rarely creates specific bonds between the lecturers and students."

> "Teaching in a lecture theatre, is hard because actually it doesn't feel natural to stand there and speak to a bunch of people for two hours where there's very little interaction"

The whole experience of co-creating a game and having adequate time to do so was viewed as a successful remedy to counter alienation in the classroom and build towards a more fruitful relationship between students and educators.

> "Communication between teachers and students made the learning process more effective, due to the constant exchange of new and unexpected ideas. Interaction with everyone is easier and less stressful compared to a lecture for example where 99% of the time you will feel shy and won't ask a question you might have. As a result of this, the designed learning material becomes more personalised and unique"

The shifting power dynamics between students and lecturers is another characteristic of co-creation processes (Bovill, Cook-Sather and Felten, 2011), the results of this study showing that co-creative GBL activities can facilitate such a shift where educators become more accessible for students to ask questions and exchange ideas with, leading to better learning.

> "There is a feeling of hierarchy in a lecture/seminar. The lecturers are untouchable because of their perceived superior knowledge. Whereas, during the workshop the relationship between students and teachers shifted to a more personal dynamic, with the active exchange of advice and ideas from both parties, which is lacking in ordinary lectures/ seminars"

**Collaboration**

Moving further from creating an environment where meaningful communication is possible, co-creative GBL activities have the potential to facilitate active collaboration between students and teachers as well as members of the student teams. The following quotes reveal feelings of ownership and pride students felt shaping their own learning and learning material as well as confirming findings of earlier research around the suitability of results of such processes:

> "I think the process of co-creation allows us to learn better as we are more involved in our own learning; we can share ideas with not only our fellow students but with our teachers. I feel this allows us to question things more openly, thus acting as a better leaning experience. The level of difficulty and complexity of the game is managed by the students themselves. It could create more stress and pressure if the students needed to play a game designed by the lecturers, because it would be very, very difficult to play"

Peer tutoring with students teaching each other even across teams was another effect of the GBL co-creation exercises where collaboration led to participants teaching each other in order to fulfil their goal and make a good learning game.

> "It's a very collaborative way to discuss project management and not just read about it in a theory book. It's teamwork, you learn from each other as you review learning content trying to it into a game. You also get some inspiration and feedback from the other team so learning happens also between groups"

Collaboration among participants was made possible by the fact that teams were working towards a specific goal, making an educational Project Management board game. Students interaction with both conceptual and physical artefacts made the exercise meaningful and more 'real' for students thus motivating them to engage with knowledge and their fellow students and teachers.

**Challenges**

Even though most of the findings indicated a positive view of participants towards the premise co-creative game- based learning activities hold for education, there were also critical points made about the process and its results. Some participants suggested for example that giving them more specific instructions would help focus on a specific knowledge area that would help them make a better game:

> "I think the game should have a more specific focus like project stakeholder management. I feel at the beginning there should be a brief saying, this is the topic that the game has to be about, and you have to make sure what you design the game with that topic in mind"

The freedom teams were given to approach the task was also considered problematic by some participants suggesting that more rigid structure to the workshop would lead to less confusion and more productivity:

> "Maybe there should be more structure in the workshops for example the 50 minutes could be organised in 5 stages of 10 minutes to apply pressure and know what needs to be done"

The final critical point about the results of the workshops revolved around the fact that students felt like the games both teams designed were engaging mainly with Project Management theory rather than practice. They felt that games could be a good chance for players to do Project Management and not just talk about Project Management:

> "Both teams made games where the only aspect of project management that we took into account are the names of project management theories and techniques but in order to actually understand them and do them in our future work we have to practice them"

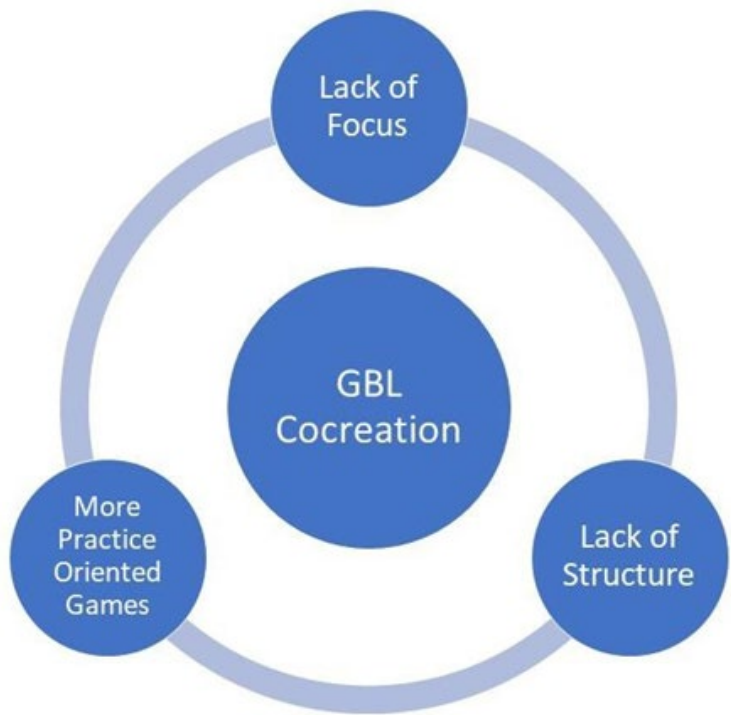

Figure 5: Areas for Improvement of Co-Creative GBL Activities

## Conclusion

This paper sought to explore the potential of co-creative GBL exercises in enhancing student engagement in Project Management education. The outcomes of the study outline some positive feedback including engagement with knowledge, assessment of knowledge, creativity, communication and collaboration. At the same time a list of critical feedback including lack of focus, lack of structure and a need for more practice- oriented games present points for improvement of future workshops. Recognising the limitations of this paper, we seek to identify future research topics that could contribute in our understanding of co-creative GBL processes and their potential to improve teaching and learning in higher education.

First, how can co-creative GBL activities be embedded in management and other curriculums? Ideas were offered by our participants that could help future researchers, such as the idea that these can be useful revision sessions where students and teachers identify what their knowledge gaps are. Implementing such activities as an official part of the curriculum and studying their effects can also further strengthen the arguments GBL and student-centred literature have been suggesting, in that increasing student participation and engagement leads to increased learning.

Second, how well do such activities scale? We facilitated two workshops with eight students, but many educators, especially in business schools, have to teach much larger cohorts even in seminars. How could such activities be adapted for a larger audience and what are the implications of that?

Third, how do co-creative GBL activities "translate" in different disciplines outside Project Management? Documenting results across multiple disciplines might yield a more pluralistic view of what can be achieved through using co-creative GBL activities as a teaching tool.

Summarising, using co-creative GBL activities in management education offer the potential to enhance the effectiveness of teaching sessions and allow for a more engaged student cohort. This paper contributes a framework outlining the merits of co-creative GBL activities and one listing potential challenges, in an effort to assist management educators create participative learning environments where students and teachers work closely together, examining and practically applying knowledge, and facilitating critical thinking geared towards solving complex real world problems.